\documentclass{article}

\usepackage{arxiv}

\usepackage[utf8]{inputenc} 
\usepackage[T1]{fontenc}    
\usepackage{hyperref}       
\usepackage{url}            
\usepackage{booktabs}       
\usepackage{amsfonts}       
\usepackage{nicefrac}       
\usepackage{microtype}      
\usepackage{lipsum}

\usepackage{graphicx}
\usepackage[caption=false,font=footnotesize,subrefformat=parens,labelformat=parens]{subfig}
\usepackage{threeparttable}
\usepackage{cite}
\usepackage{amsmath,amssymb,amsfonts}
\usepackage{algorithmic}
\usepackage{textcomp}
\usepackage{multirow}
\usepackage{soul}

\title{Automatic Qubit Characterization and Gate Optimization with \textit {QubiC}}

\author{
 Yilun Xu, Gang Huang\thanks{Corresponding author. Email: ghuang@lbl.gov}, Jan Balewski, Alexis Morvan, Kasra Nowrouzi, David I. Santiago \\
 Lawrence Berkeley National Laboratory, Berkeley, CA 94720, USA\\
 \And
 Ravi K. Naik, Brad Mitchell, Irfan Siddiqi\\
 University of California at Berkeley, Berkeley, CA 94720, USA\\
}

\begin{document}
\maketitle

\begin{abstract}
As the size and complexity of a quantum computer increases, quantum bit (qubit) characterization and gate optimization become complex and time-consuming tasks.
Current calibration techniques require complicated and verbose measurements to tune up qubits and gates, which cannot easily expand to the large-scale quantum systems.
We develop a concise and automatic calibration protocol to characterize qubits and optimize gates using {\it QubiC}, which is an open source FPGA (field-programmable gate array) based control and measurement system for superconducting quantum information processors.
We propose mutli-dimensional loss-based optimization of single-qubit gates and full XY-plane measurement method for the two-qubit CNOT gate calibration.
We demonstrate the {\it QubiC} automatic calibration protocols are capable of delivering high-fidelity gates on the state-of-the-art transmon-type processor operating at the Advanced Quantum Testbed at Lawrence Berkeley National Laboratory.
The single-qubit and two-qubit Clifford gate infidelities measured by randomized benchmarking are of $4.9(1.1) \times 10^{-4}$ and $1.4(3) \times 10^{-2}$, respectively.
\end{abstract}


\section{Introduction}
Quantum computers, which harness and exploit the laws of quantum mechanics to process information, have the potential to revolutionize computation by making  solvable certain types of classically intractable problems \cite{arute2019quantum,preskill2018quantum}. 
In the near term noisy intermediate-scale quantum (NISQ) computing era, a lot of quantum algorithms have been developed on a broad range of applications \cite{montanaro2016quantum,suau2021practical,ushijima2021multilevel}.
Quantum algorithms are most commonly described by quantum circuits, which consist of quantum gates performed on one or more quantum bits (qubits).
The calibration of a quantum processor is the process of finding optimal control parameters to construct quantum gates to steer the evolution of quantum systems.
Since many of the most severe errors emerge from imprecise calibration and system drift, qubit characterization and gate optimization are essential for successful to quantum computations \cite{kelly2018physical,klimov2020snake}.

Initial setup of a physical multi-qubit system is a complex procedure which involves multiple instruments and multiple optimizations \cite{xu2020automatic,xu2021automatic}. 
As the size and complexity of the quantum system increases, the manual qubit characterization and gate optimization will be a time-consuming and not extensible task. 
The physical qubit must be carefully calibrated routinely because quantum information processors are sensitive to the environment \cite{proctor2020detecting} and the control hardware can have slow drift with time \cite{corcoles2019challenges} so as to impact the gate fidelity. 
However, existing calibration techniques \cite{rudinger2017experimental,rol2017restless,patterson2019calibration,white2021performance} require complicated and verbose measurements to tune up multiple parameters of each gate independently, which cannot easily expand to the large-scale quantum systems.
Furthermore, a concise calibration protocol has the possibility to be fully automated, which is a very desired feature for the multi-qubit system.

Here we present an automatic qubit characterization and gate optimization method with {\it QubiC} (Qubit Control) system to tackle the calibration challenges and keep pace with rapidly evolving classical control requirements. 
{\it QubiC} system is an open source, FPGA (field-programmable gate array) based, control and measurement system for superconducting quantum information processors \cite{xu2020qubic, xu2020rf}. 
The system consists of electronics hardware, FPGA gateware, and engineering software.
Leveraging the state-of-the-art FPGA technology, {\it QubiC} provides fully parametric waveform generation, analog response acquisition and manipulation, and classical signal post-processing.
{\it QubiC} allows researchers to control all levels of the software stack, which enables the execution of a broader class of computation experiments while also facilitating the implementation co-design at each level of the control stack in next generation systems.

\section{Single-qubit Characterization and Gate Optimization}
The qubit bias point and the waveform of pulses driving gates on a qubit require per-qubit calibration. The calibration values depend on the various details of the chip manufacturing process, the readout and control circuitry, and  in-fridge connectivity. Every qubit requires an initial detailed calibration, supplemented by a frequent re-calibration process. 
An arbitrary single-qubit $U_3$-gate, parametrized by 3 arbitrary angles, can always be decomposed into a sequence of two rotations of the quantum state
 by 90 deg along the X-axis and 3 virtual-Z gates with 3 arbitrary angles~\cite{PhysRevA.96.022330}. Since the virtual-Z gate is applied in software it does not require a calibration.
Therefore, obtaining a high fidelity  X(90) gate
~\footnote{In the Bloch sphere representation the X(90) and X(180) gates rotate qubit state around the X-axis by 90 and 180 deg, respectively.}
is the primary objective of the single-qubit calibration, allowing for an arbitrary $U_3$-gate to be executed on this qubit. 

We are focusing on superconducting transmon-type qubits~\cite{PhysRevA.76.042319},  readout via  coupling capacitance, and controlled by a 5-6~GHz microwave pulses. The 4 initial calibration parameters, aka qubit bias, are:  qubit frequency, readout resonator frequency, qubit drive amplitude, and readout drive amplitude. A typical  qubit driving pulse induces Rabi oscillations  between the ground and the 1st excited state. The functional form of the Rabi oscillations is expected to follow the cosine shape with the amplitude decaying exponentially
\begin{eqnarray}
\label{eq:sinexp}
P(t)&=& C+A\exp(-\tau t)\sin(2\pi f t +\phi_0)
\end{eqnarray}
where $t$ is the length of the the Rabi waveform and  $C, A, \tau,f,\phi_0$ parameters are fitted to the data.

{\it QubiC} contains the software and established procedures allowing for the initial calibration and re-calibration of the X(90) and X(180) gates.

\begin{figure}[!ht]
\centering
\includegraphics[width=.48\linewidth]{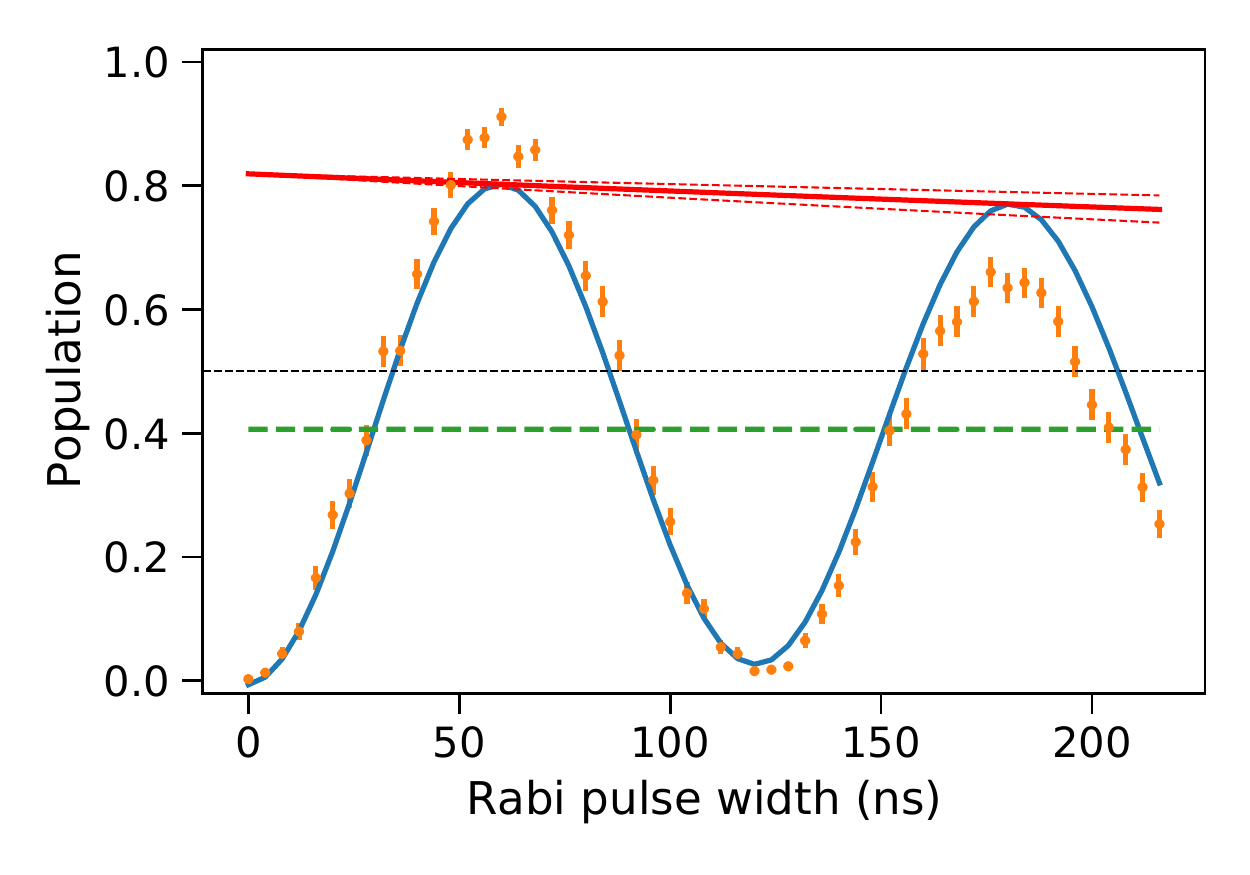}~~~~~
\includegraphics[width=.48\linewidth]{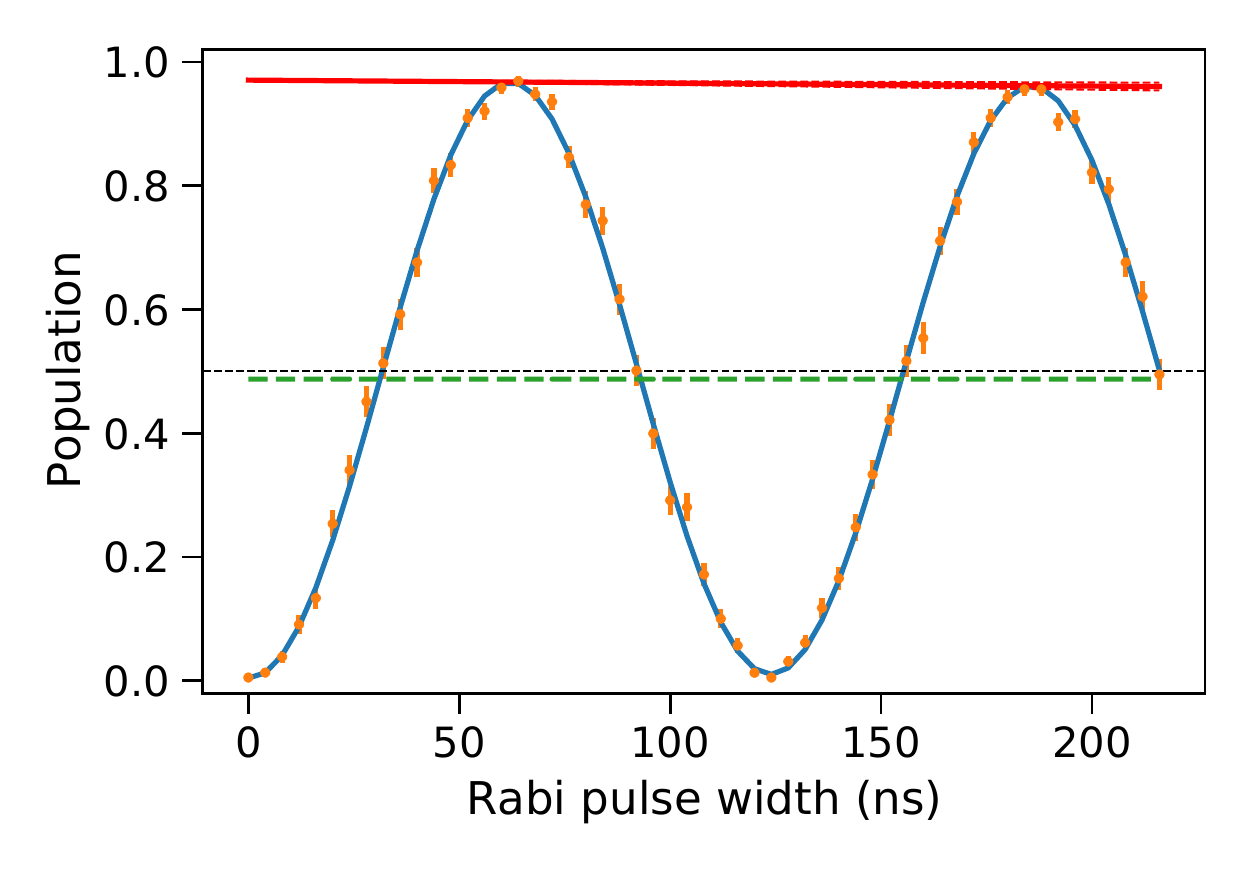}
\caption{ Rabi oscillations induced on a qubit fitted with shape defined in eq.~\ref{eq:sinexp}.  Left: A crude initial calibration of the Rabi waveform results with a low amplitude, quickly decaying  Rabi oscillations. The poor agreement of measured data (+) with the fitted shape (line)  results with a large $\chi^2$. Right: The calibration from {\it autoRabi} optimizer yields much better signal.  The maximal observed population is not reaching 1.0 because of the intrinsic hardware limitations. The  readout  correction has not been applied.}
\label{fig:rabi}
\end{figure}

The initial survey of the response of each qubit is done by sequentially  exciting  the resonator and next, by inducing a mensurable Rabi oscillations of the qubit, as shown in Fig.~\ref{fig:rabi}a. The resonator readout amplitude is chosen based on linearity of the qubit response and is set at the possible high value to achieve good separation of the 0- and 1-states in the IQ-plane but not too high to avoid cross-talk between the qubits or  populating the the 2nd quantum state~\cite{Sank_2016}.

\subsection{Automatic X(90) Gate Calibration}
The qubit-resonator system responds to the qubit and the resonator driving pulses in a complex, non-linear fashion. Often, the  3 parameters: qubit frequency ($f_q)$,  Rabi drive amplitude ($A_R$), and  readout frequency ($f_r)$ are optimized sequentially, what prevents adjustment of say $f_r$ while optimizing for $f_q$, {\it etc}. 

We develop the
{\it autoRabi} procedure which aims to calibrate the X(90) gate by  simultaneous adjustment of all 3 parameters. {\it autoRabi} minimizes the $\chi^2$-type loss function ($\mathcal{L}_{tot} $) over the vector of the unit-less quantities of concern ($\overrightarrow{\Delta}$)
\begin{eqnarray}
\label{eq:arl}
\mathcal{L}_{tot}&\!=& \mathcal{L}_{F} + \mathcal{L}_{AC}+\mathcal{L}_{T}+ \mathcal{L}_{BIC}\\
&\!=&\sum \Delta_i^2\\
\overrightarrow{\Delta}&\!=&  \!\!\left[\! ~\sqrt{\chi^2_{NDF}},  ~~\frac{|A|-0.5}{0.03} , ~~\frac{C-0.5}{0.05}, ~~\frac{\frac{T_{X(90)}}{ns}-32}{4.0},  
~~\frac{\sigma\left(\delta_{BIC}(1)\right)}{0.5}, 
~~\frac{\sigma\left(\delta_{BIC}(3)\right)}{0.5}, 
~~\frac{\sigma\left(\delta_{BIC}(4)\right)}{0.5} 
~\!\right]
\end{eqnarray}
The total loss $\mathcal{L}_{tot} $ is composed of 4 unit-less  terms:
\begin{itemize}
    \item  $\mathcal{L}_{F} $ term assures the measured population matches the expected Rabi shape (eq.~\ref{eq:sinexp}). If the fit is poor, as in Fig.~\ref{fig:rabi}a,  this term is large and dominates the loss.
    
    \item$\mathcal{L}_{AC}$ expresses our preference for maximizing the Rabi amplitude (A) and the equal average population (C) of 0- and 1-states. The normalization constants of 0.03 and 0.05  are found empirically and are meant to reflect a subjective trade-off between the goodness of the fit and the magnitude of the Rabi contrast.
    
    \item $\mathcal{L}_{T}$ enforces the quantum state is rotated by 90 deg by the Rabi waveform of the length of 32~ns.
The choice of normalization  results with a 4 ns deviation of the Rabi period (equivalent to 360 deg rotation) is as important as a change of the $\chi^2_{NDF}$ by 1 unit.

    \item $\mathcal{L}_{BIC}$ ensures the hypothesis of the existence of 2-and-only-2 clusters  in the IQ-plain is the most probable hypothesis. It is computed from Bayesian Information Criterion (BIC) \cite{wiki:bic}.
\end{itemize}

\begin{figure}[!ht]
\centering
\includegraphics[width=.32\linewidth]{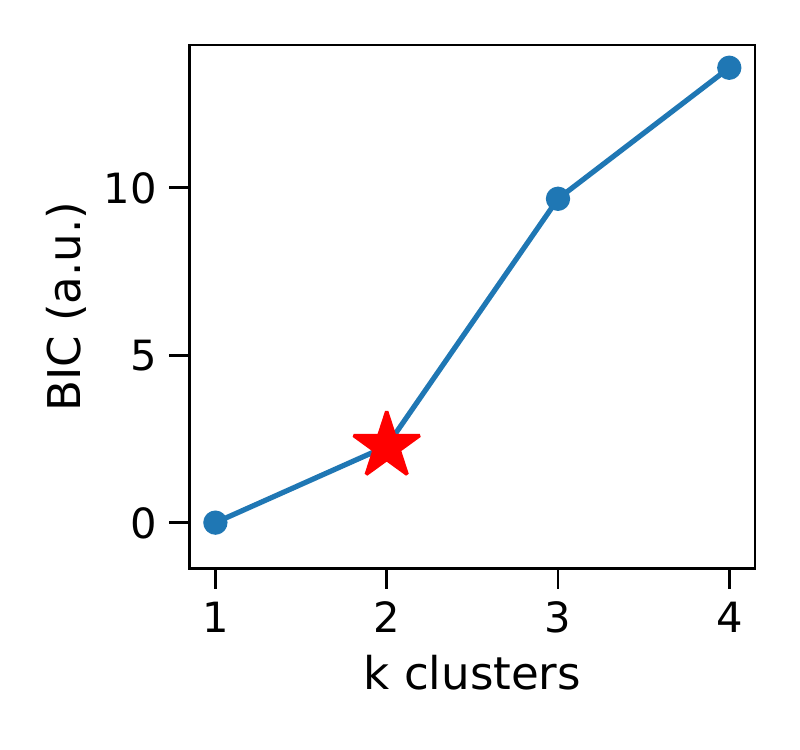}
\includegraphics[width=.32\linewidth]{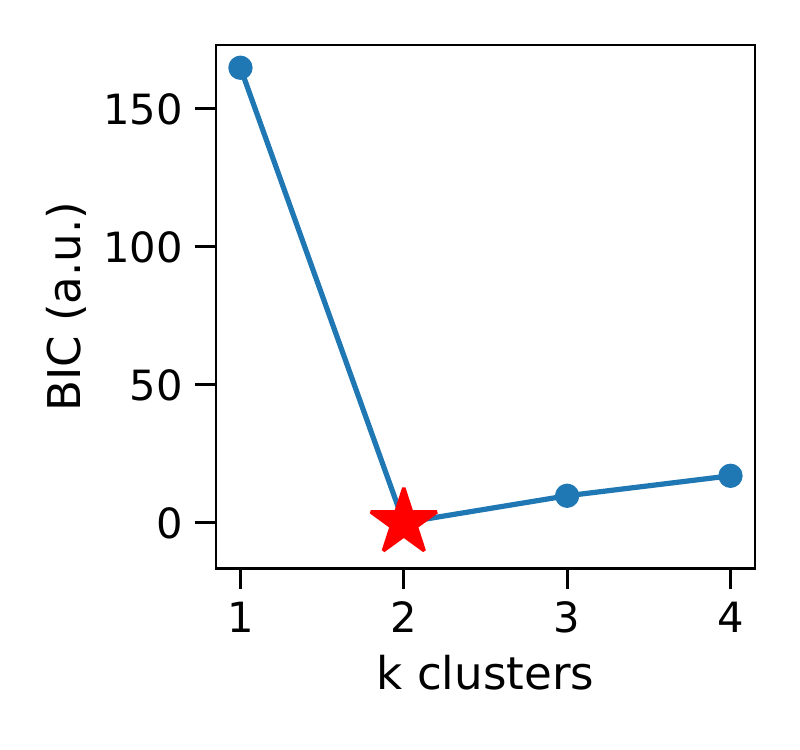}
\includegraphics[width=.32\linewidth]{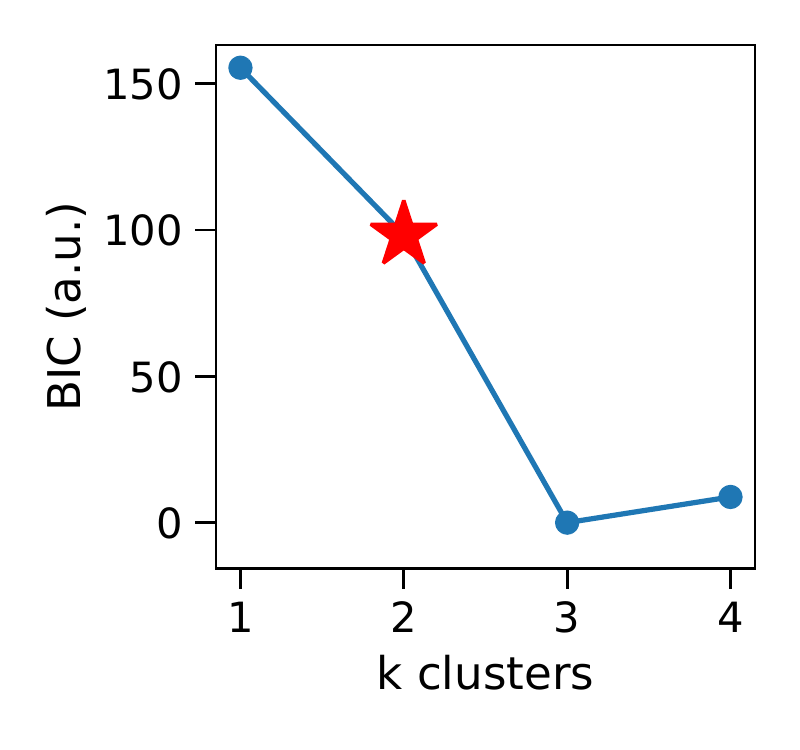}
\\
\includegraphics[width=.32\linewidth]{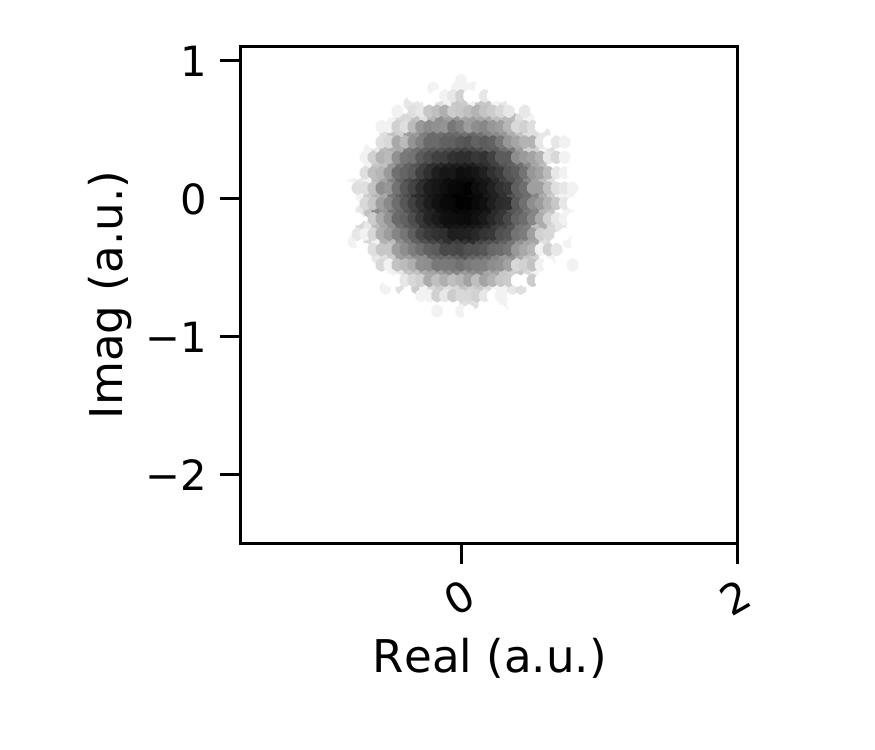}
\includegraphics[width=.32\linewidth]{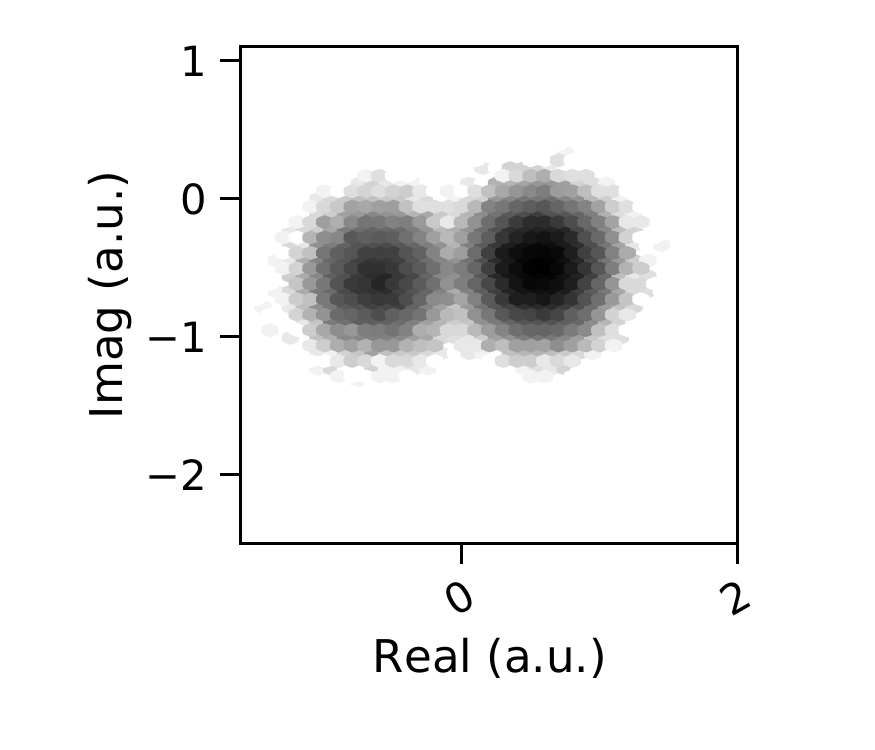}
\includegraphics[width=.32\linewidth]{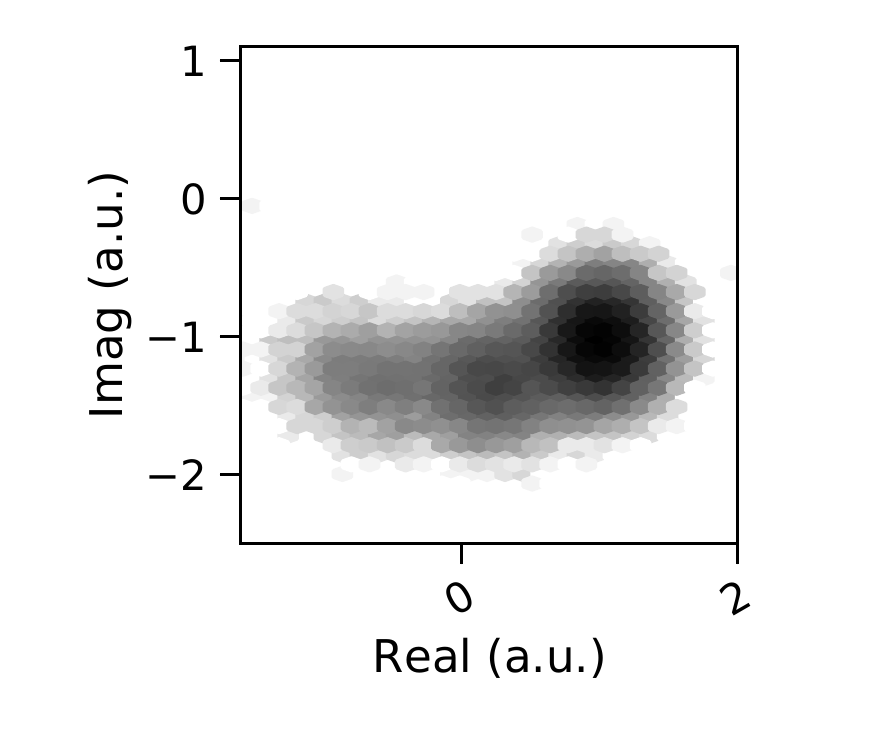}
\caption{ Top: Bayesian Information Criterion $BIC_k$ as function of assumed number of clusters (k) in the IQ-plane.  3 columns  correspond to 3 types of data-sets with the true 1, 2, and 3 clusters, respectively.  BIC has the minimum for the correct hypothesis of number of clusters. Bottom: raw IQ-pairs data-sets used as  inputs, the log of counts is represented by the gray scale.}
\label{fig:bic}
\end{figure}

The $BIC_k$ is a criterion for model selection among a finite set of models; the model with the lowest BIC is the most likely. In our case we vary the number of clusters (k), as shown in Fig.~\ref{fig:bic}(top). Since the magnitude of BIC is very large and also depends  on the sample size a regularization is needed before it can be added to the  total loss $\mathcal{L}_{tot}$ in a robust fashion.
We decided to combine the 3 BIC-based constraints 
\begin{eqnarray}
\label{eq:clust}
BIC_2 < BIC_k ~~~~~~ \textrm{for} ~k \in [1,3,4]
\end{eqnarray}
using the sigmoid function $\sigma(.)$   applied on the scaled difference of BICs 
\begin{eqnarray}
\sigma(x)&=&\frac{1}{1+e^{-x}}\\
\delta_{BIC}(k)&=& ( BIC_k - BIC_2)/10
\end{eqnarray} where the  factor of 10 is chosen to attenuate the changes of BIC with k.  Combining all the above leads to $\mathcal{L}_{BIC}$ which vanishes if all constraints in eq.~\ref{eq:clust} are meat,  otherwise $\mathcal{L}_{BIC}$  never exceeds the 4 units of $\chi^2_{NDF}$  to not overwhelm the  $\mathcal{L}_{tot} $. 

To summarize, this definition of the total loss  $\mathcal{L}_{tot} $ simultaneously pushes the qubit bias point toward (i)   fit to the Rabi spectrum is good ~($\chi^2_{NDF}\sim 1)$, (ii)  Rabi contrast  is good($A\sim C\sim 0.5$), (iii) the Rabi period is on target,  (iv)  the 2-and-only-2 clusters are most likely in the measured IQ-pairs distribution.

The muti-dimensional optimization procedure is using COBYLA 
\footnote{Constrained Optimization By Linear Approximation }
minimizer~\cite{powell_1998}. It requires  a prior knowledge of the approximate qubit bias point at which  a recognizable Rabi oscillations are observed, as shown in Fig.~\ref{fig:rabi}a. We have   bracket of the  range of changes of  the qubit  and readout   frequencies ($f_q, f_r$) at 2 MHz,  and the Rabi drive amplitude change ($A_R$) at 0.3 of the full scale.  For each iteration of COBYLA the following steps are executed:
\begin{enumerate}
    \item COBYLA proposes a new set of values for: $f_q, f_r, A_R$,
    \item {\it QubiC} generates appropriate wave-forms, drives the qubit for the fixed number of shots, records and saves to discs the raw IQ-pairs,
    \item 4 GMM~\footnote{Gaussian Mixture Model} ~\cite{wiki:gmm} fits are executed, assuming k=[1..4] clusters, needed to collect $\delta_{BIC}(k)$ information,
    \item GMM(k=2) is used to digitize the IQ-pairs into 0 or 1 bits and the probability of the 1-state is computed as a function of the length of the Rabi waveform,
    \item Rabi oscillation parameters (A,C) and the $\chi^2_{NDF}$ are extracted from the fit to the measured data,
    \item  value of loss $\mathcal{L}_{tot} $ is computed based on eq.~\ref{eq:arl} and feed back to the next iteration of COBYLA.
\end{enumerate}
Typically, {\it autoRabi} converges in about  half an hour, after   40 iterations, each requiring {\it QubiC} to take 400 shots for 50 widths of the Rabi waveform. A typical improvement in the Rabi contrast is illustrated in  Fig.~\ref{fig:rabi}b.

\subsection{Fine Tuning of X(90) and X(180) Gates}
\begin{figure}[!ht]
\centering
\includegraphics[width=.48\linewidth]{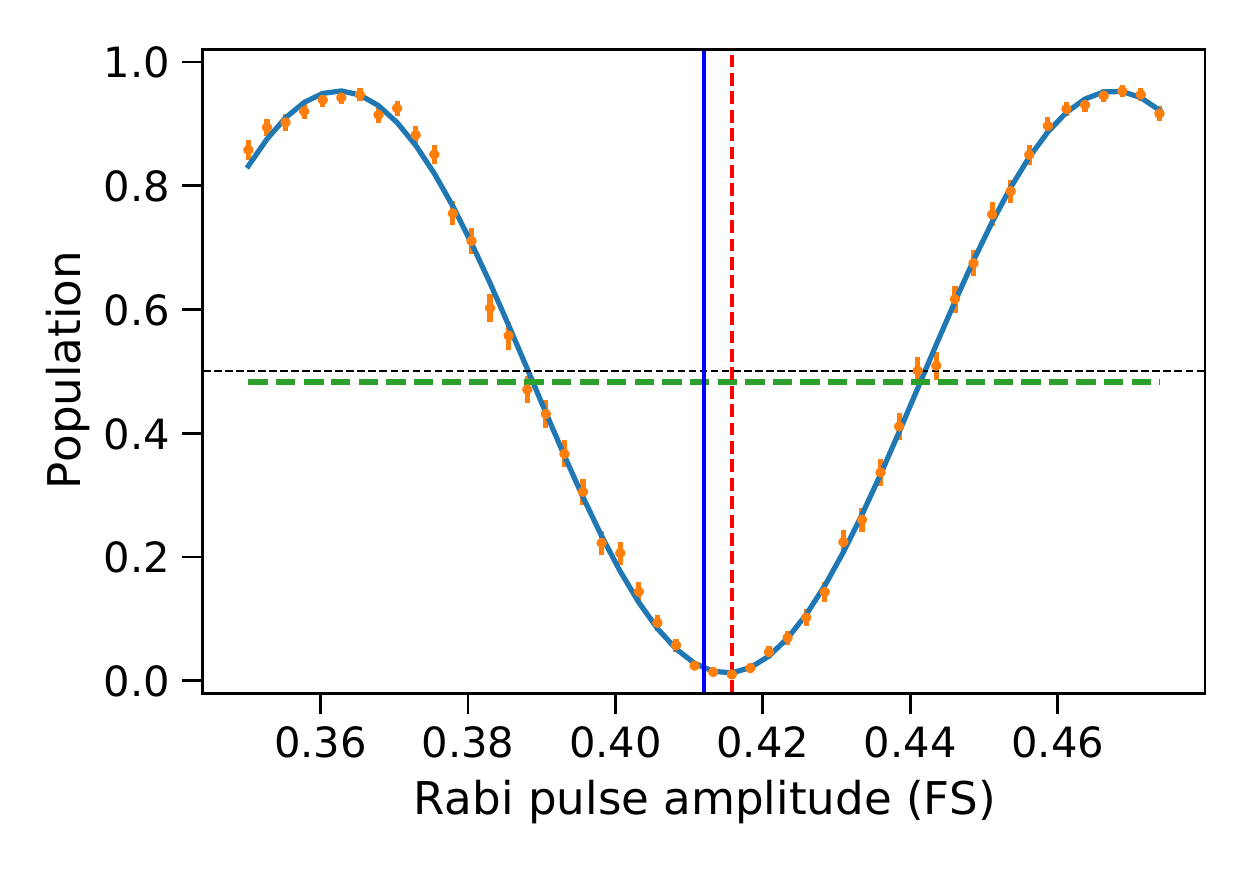}~~~~~ 
\includegraphics[width=.48\linewidth]{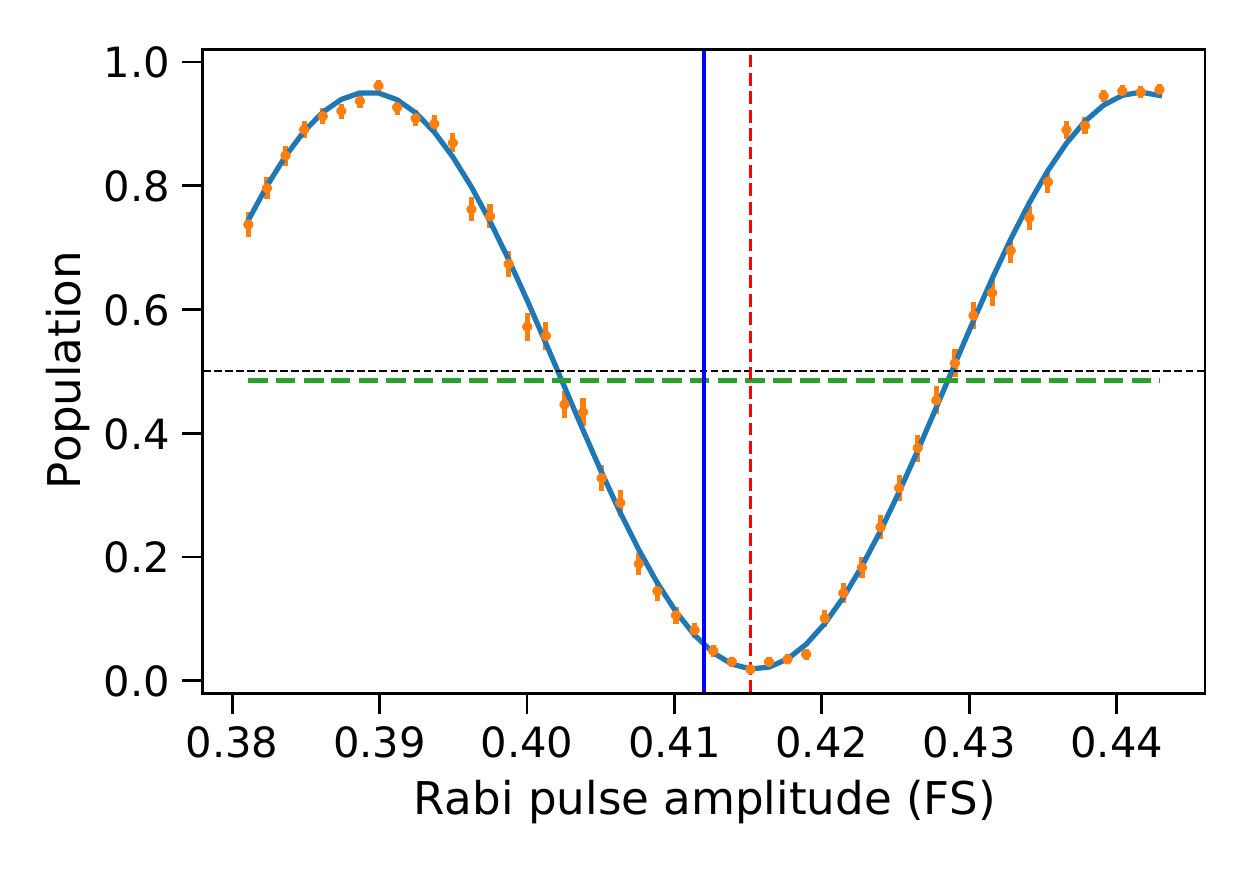}
\caption{ Stacking of 16  X(90) gates (left) and 16  X(180) (right) gates applied on the 0-state qubit is equivalent to an identity-gate and allows for the fine-tuning of the waveform amplitude. The blue (solid) line is the initial value from {\it autoRabi}, the red (dashed) line is the optimal amplitude.  }
\label{fig:stackRabi}
\end{figure}

The  $f_q, f_r, A_R$ values found by the {\it autoRabi} define the X(90) gate with a precision of 1-2 deg of the state rotation. To achieve a better accuracy we stack large numbers of the same gate and adjust the amplitude $A_R$, while  keeping the duration of the gate constant
~\footnote{ Due to the {\it QubiC} hardware limitation any waveform length must be a multiplicity of 4 ns, tied to the fixed frequency of the FPGA clock }.
We prepare the initial qubit state to be 0 and  stack the X(90) gate N*4 times to achieve a nominal identity gate.  We sweep the $A_R$ values in the vicinity of the  value from {\it autoRabi}  and fit the measured population probability with the sin-function. The optimal gate amplitude is found as  the {\it arg minimum}  of the fit,  shown as vertical red (dashed) line in Fig.~\ref{fig:stackRabi}a. The optimal amplitudes for X(90) and X(180) differ a little because of the difference in the ratio of the length of the rising edge to the gate duration.

The single-gate calibration procedures discussed in this section can be performed using {\it QubiC} software stack and allow for implementation of an arbitrary $U_3$-gate. {\it QubiC} software also contains tools to calibrate the DRAG  corrections~\cite{motzoi2009simple}, fine-tune the qubit drive frequency with $T_2$~Ramsey  measurement, as well as  measure $T_1$ and $T_2$~spin-echo. 

The ultimate test of the fidelity of quantum gates is provided by the randomized benchmarking protocol and is discussed in the following section~\ref{sec:rb}.

\section{Two-qubit Gate (CNOT) Calibration}

\subsection{Full Entanglement of Cross Resonance}
\begin{figure}[t!]
\centering
\includegraphics[width=0.6\linewidth]{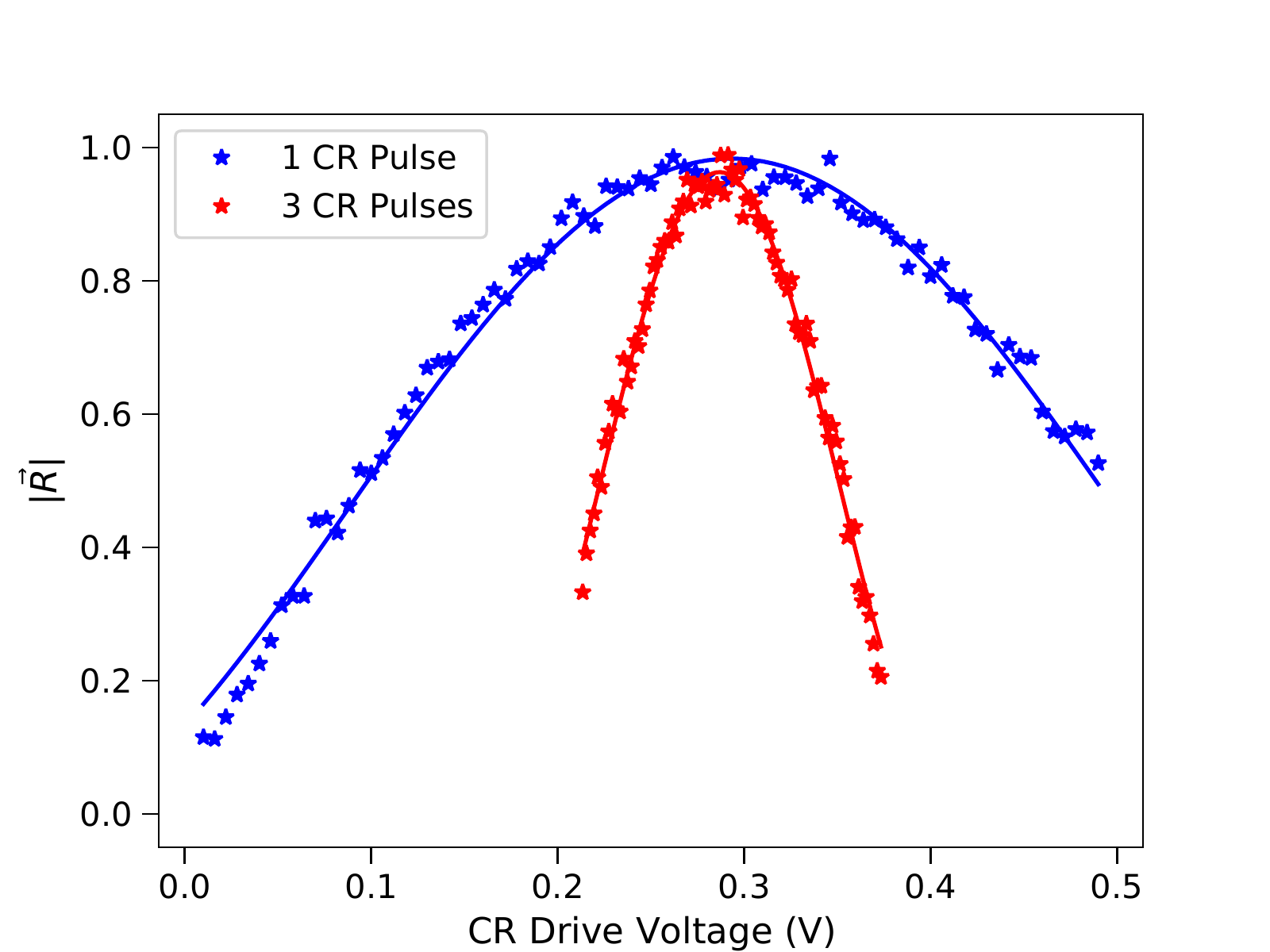}
\caption{CR measurement to search for the full entanglement. One and three CR pulses are employed for coarse and fine identification respectively. The optimal value is found as {\it arg maximum} of the parabola fit.}
\label{fig:cr}
\end{figure}

Cross resonance (CR) is one of the methods to realize the two-qubit entanglement gate for the superconducting fixed-frequency transmon qubits \cite{sheldon2016procedure}. 
We optimize the CR pulse of the cosine edge envelope with fixed ramp length to achieve maximal entanglement on the control and target qubits. 
To tune up a CR gate, we prepare the initial states of both qubits to be |0>, and then apply the CR gate, followed by the target qubit state projection onto X, Y, and Z axis.
The same process is repeated for the control qubit in |1>. 
We fix the CR pulse length and sweep the CR pulse amplitude, resulting with 6 independent measurements which are combined into one value:
the Bloch vector length $|\vec{R}|$ of the target qubit:
\begin{equation}
  \label{eq3}
  \lvert \vec{R} \rvert=\frac{1}{2}\sqrt{(\mathrm{X}_0-\mathrm{X}_1)^2+(\mathrm{Y}_0-\mathrm{Y}_1)^2+(\mathrm{Z}_0-\mathrm{Z}_1)^2},
\end{equation}
where X, Y, Z are the expectation values of the corresponding components of the target qubit, while the subscript 0, 1 are preparation states of the control qubit.
The goal of the CR pulse amplitude sweeping is to achieve the full entanglement, where $|\vec{R}|$ is maximized.
We can stack multiple CR pulses to make the $|\vec{R}|$ curve sharper to find the maximum more accurately.
As shown in Fig.~\ref{fig:cr}, we implement one CR pulse for coarse identification and stacked three CR pulses for precise determination in the real measurement.

\subsection{Full XY-plane Measurement and Fitting of CNOT}
The CR pulse of particular length and amplitude can set two qubits in the fully entangled state, but typically additional single-qubit rotations are induced on each of the participating qubits.
We need to attach a set of appropriate single-qubit gates before and after the CR pulse to obtain the real CNOT gate. 
It is common practice to project the qubit state to X/Y/Z axis for diagnosis.
This is enough to evaluate if the state is good or not, but it would be good to gather more information about the state to help the optimization.
Here we develop a method called full XY-plane measurement, which projects both qubits states onto the measurement axis at the same angle in the XY-plane, sweeps the angle, and extracts the CNOT parameters from the curve fitting.

\begin{figure}[t!]
\centering
\includegraphics[width=1.0\linewidth]{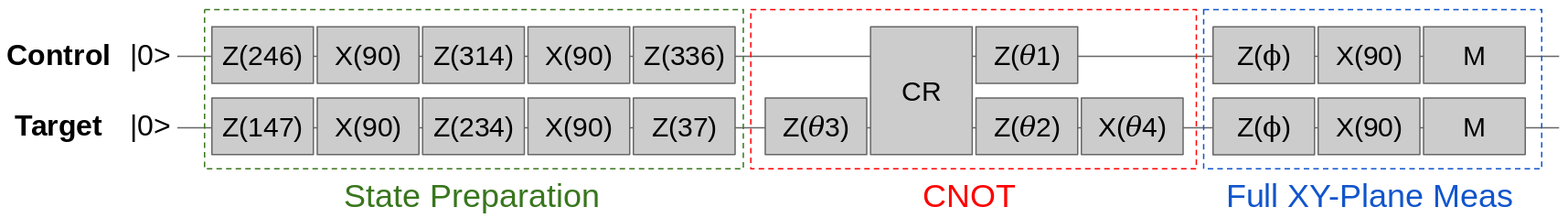}
\caption{Full XY-plane measurement schematic. Circuits are operated on the control qubit line and the target qubit line. The gate rotation angle is in the unit of degree. M represents a measurement.}
\label{fig:fullxy_meas_shcmatic}
\end{figure}

\begin{figure}[t!]
\centering
\includegraphics[width=0.6\linewidth]{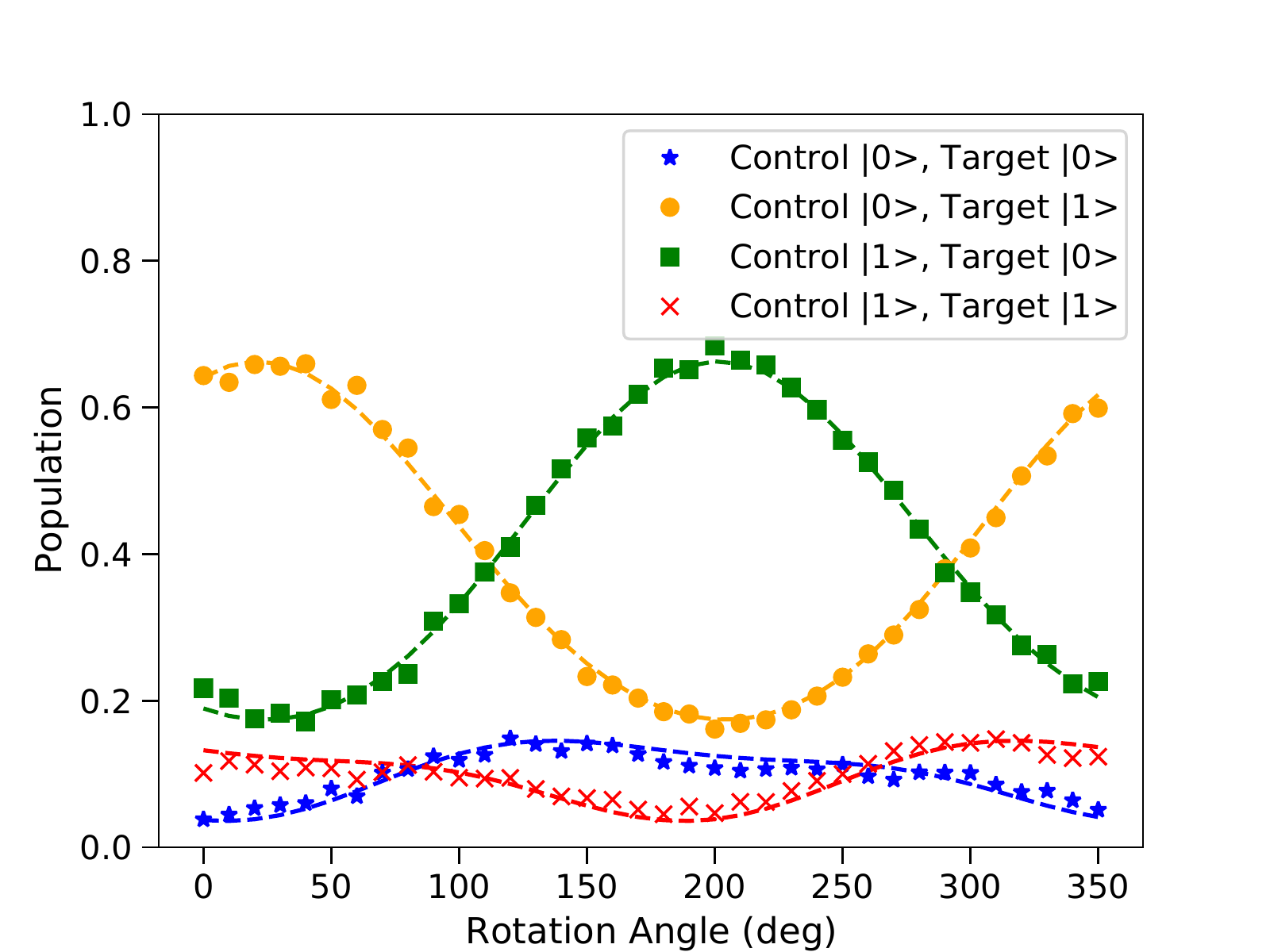}
\caption{Full XY-plane measurement (star points) and fitting (dashed lines) results. Four different colors represent four basis states $|00>$, $|01>$, $|10>$, $|11>$ of control and target qubits. We can see four unique patterns as a function of rotation angle $\phi$ among the four basis states, which can be distinguished by curve fitting. Each data point is an average of $\sim$2000 shots. The full XY-plane measurement and fitting only take $\sim$1.5 minutes.}
\label{fig:fullxy_meas_fit}
\end{figure}

As shown in Fig.~\ref{fig:fullxy_meas_shcmatic}, firstly we need to prepare states for both control and target qubits. 
In general, the qubit state can be randomly chosen, as long as it is not sitting at the south pole or north pole on the Bloch sphere. 
A CNOT gate with zero initial values is employed after the state preparation.
We then apply a virtual $Z(\phi)$ gate on a degenerate prepared state for both the control and target qubits, followed by an X(90) gate for rotation, and we measure.
Scanning the Z phase ($\phi$) will project both states to any angle on the XY-plane.
By scanning over the full XY-plane, we spread the noise among different angles, enabling us to extract the CNOT parameters from curve fitting by only executing one set of measurement, as shown in Fig.~\ref{fig:fullxy_meas_fit}. 

We assume, the CNOT gate relates to CR gate as follows:
\begin{equation}
\label{eq:cr2cnot}
CNOT=IX(\theta_4)\cdot ZZ(\theta_1,\theta_2)\cdot CR\cdot IZ(\theta_3),
\end{equation}
where the CR represents the unitary for the CR pulse.
If we describe the IX, ZZ, and IZ terms in the matrix form, we will have
\begin{equation}
IX(\theta_4)=I \otimes R_X(\theta_4)=\begin{pmatrix} 1 & 0  \\ 0 & 1 \\ \end{pmatrix} \otimes \begin{pmatrix} \mathrm{cos}\frac{\theta_4}{2} & -i\mathrm{sin}\frac{\theta_4}{2}  \\ -i\mathrm{sin}\frac{\theta_4}{2} & \mathrm{cos}\frac{\theta_4}{2} \\ \end{pmatrix},
\end{equation}
\begin{equation}
ZZ(\theta_1,\theta_2)=R_Z(\theta_1) \otimes R_Z(\theta_2)=\begin{pmatrix} e^{-i\frac{\theta_1}{2}} & 0 \\ 0 & e^{i\frac{\theta_1}{2}} \\ \end{pmatrix}  \otimes \begin{pmatrix} e^{-i\frac{\theta_2}{2}} & 0 \\ 0 & e^{i\frac{\theta_2}{2}} \\ \end{pmatrix},
\end{equation}
\begin{equation}
IZ(\theta_3)=I \otimes R_Z(\theta_3)=\begin{pmatrix} 1 & 0 \\ 0 & 1 \\ \end{pmatrix}  \otimes \begin{pmatrix} e^{-i\frac{\theta_3}{2}} & 0 \\ 0 & e^{i\frac{\theta_3}{2}} \\ \end{pmatrix},
\end{equation}
where $\theta_1$, $\theta_2$, $\theta_3$, and $\theta_4$, are ZI phase, IZ phase, ZX phase, and IX phase respectively. 
These four single-qubit parameters around the CR pulse need to be calculated from the curve fitting.
$R_X$, $R_Z$ represent the single-qubit rotation about the X, Z axis, while I is the identity gate.
The symbol $\otimes$ denotes the tensor product between two single-qubit rotation matrices.
Since the CR pulse starting phase is not accumulated in the phase calculation, we add a constraint $\theta_2+\theta_3=2\pi$ in order to use $\theta_2$ to cancel out $\theta_3$.

\begin{figure}[t!]
\centering
\includegraphics[width=0.5\linewidth]{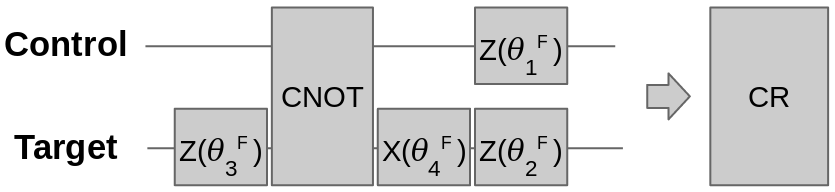}
\caption{Full XY-plane fitting schematic.}
\label{fig:fullxy_fit_shcmatic}
\end{figure}

Now we take the inverse step of measurement schematic, adjusting the $\theta_1$, $\theta_2$, $\theta_3$, $\theta_4$ to find the curve to match the measurement as shown in Fig.~\ref{fig:fullxy_fit_shcmatic}. 
In the real measurement, we start with all the parameters ($\theta_1$, $\theta_2$, $\theta_3$, $\theta_4$) equal to 0, as if the CR is a CNOT. 
The fitting of the parameters will tell us how far away the CR is from the real CNOT.
If we describe the CR matrix as  
\begin{equation}
\label{eq:cnot2cr}
CR=ZZ(\theta_1^F,\theta_2^F)\cdot IX(\theta_4^F)\cdot CNOT\cdot IZ(\theta_3^F),
\end{equation}
the CNOT matrix can be derived as
\begin{equation}
CNOT=IX^{-1}(\theta_4^F)\cdot ZZ^{-1}(\theta_1^F,\theta_2^F)\cdot CR\cdot IZ^{-1}(\theta_3^F)
=IX(-\theta_4^F)\cdot ZZ(-\theta_1^F,-\theta_2^F)\cdot CR\cdot IZ(-\theta_3^F),
\end{equation}
where the superscript $F$ represents the fitting parameters.
The negative sign before the fitting parameters means that the final $\theta$ parameter will be the difference between the initial value and the fitted value, that is $\theta=\theta_{\mathrm{init}}-\theta_{\mathrm{fit}}$.
As shown in Fig.~\ref{fig:fullxy_meas_fit}, the measurement result agrees well with the matrix calculation, so that the CNOT parameters can be extracted from the curve fitting by only implementing one set of measurement.
It is noted that the full XY-plane measurement method will yield two sets of parameters for the CNOT gate, with a global phase of $\pi/2$ between them:
\begin{equation}
CNOT=IX(\theta_4)\cdot ZZ(\theta_1,\theta_2)\cdot CR\cdot IZ(\theta_3),
\end{equation}
\begin{equation}
e^{i\frac{\pi}{2}}CNOT=IX(-\theta_4)\cdot ZZ(\theta_1-\pi,\theta_2-\pi)\cdot CR\cdot IZ(\theta_3+\pi).
\end{equation}


\section{Validation by Randomized Benchmarking}
\label{sec:rb}
\begin{figure}[t!]
\centering
\subfloat[Single-qubit Streamlined RB.]{\includegraphics[width=0.33\linewidth]{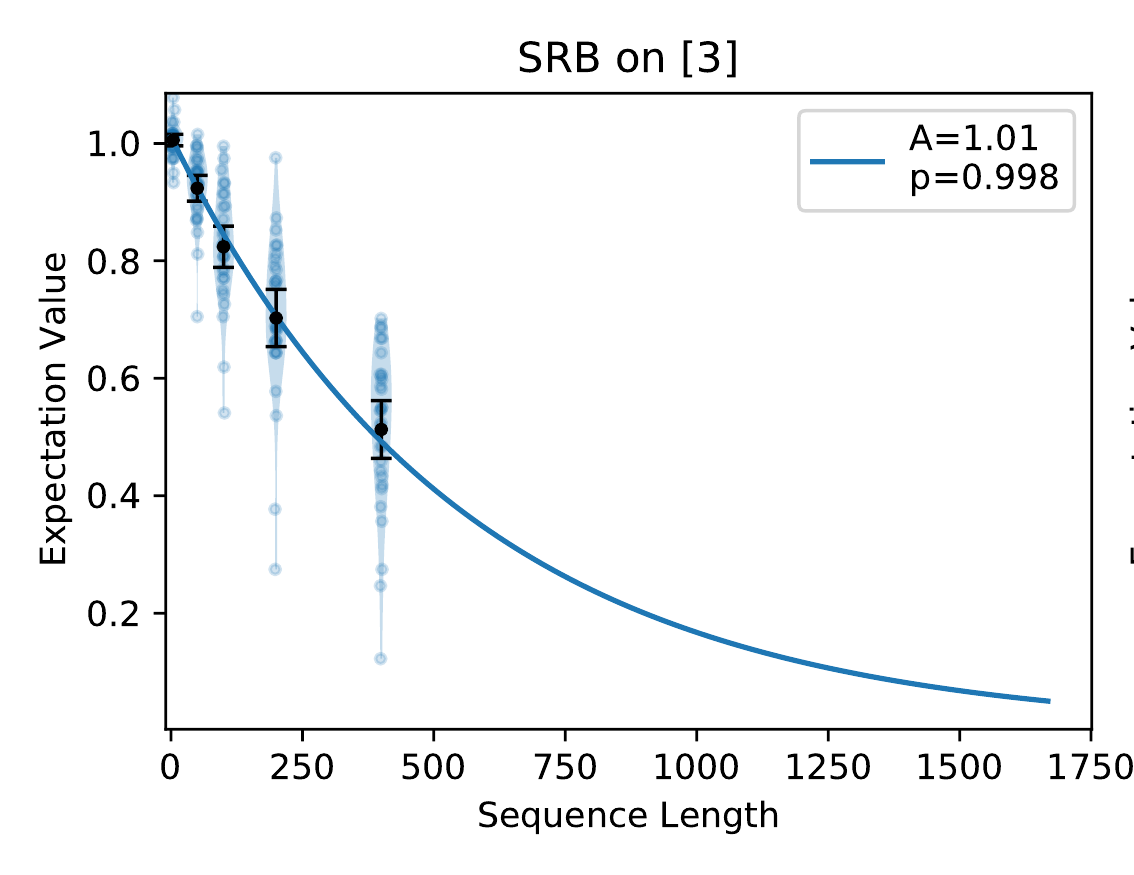}
\label{fig:1qsrb}}
\subfloat[Single-qubit Interleaved RB.]{\includegraphics[width=0.33\linewidth]{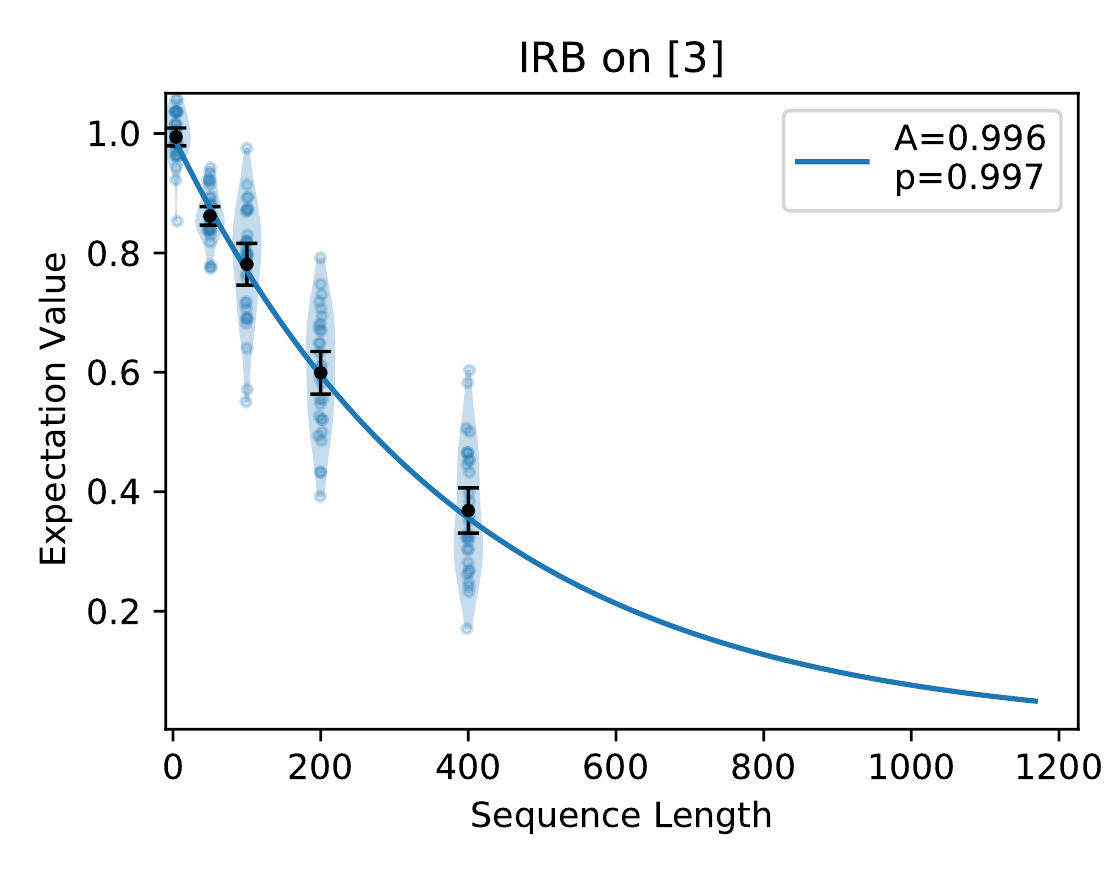}
\label{fig:1qirb}}
\subfloat[Single-qubit Extended RB.]{\includegraphics[width=0.33\linewidth]{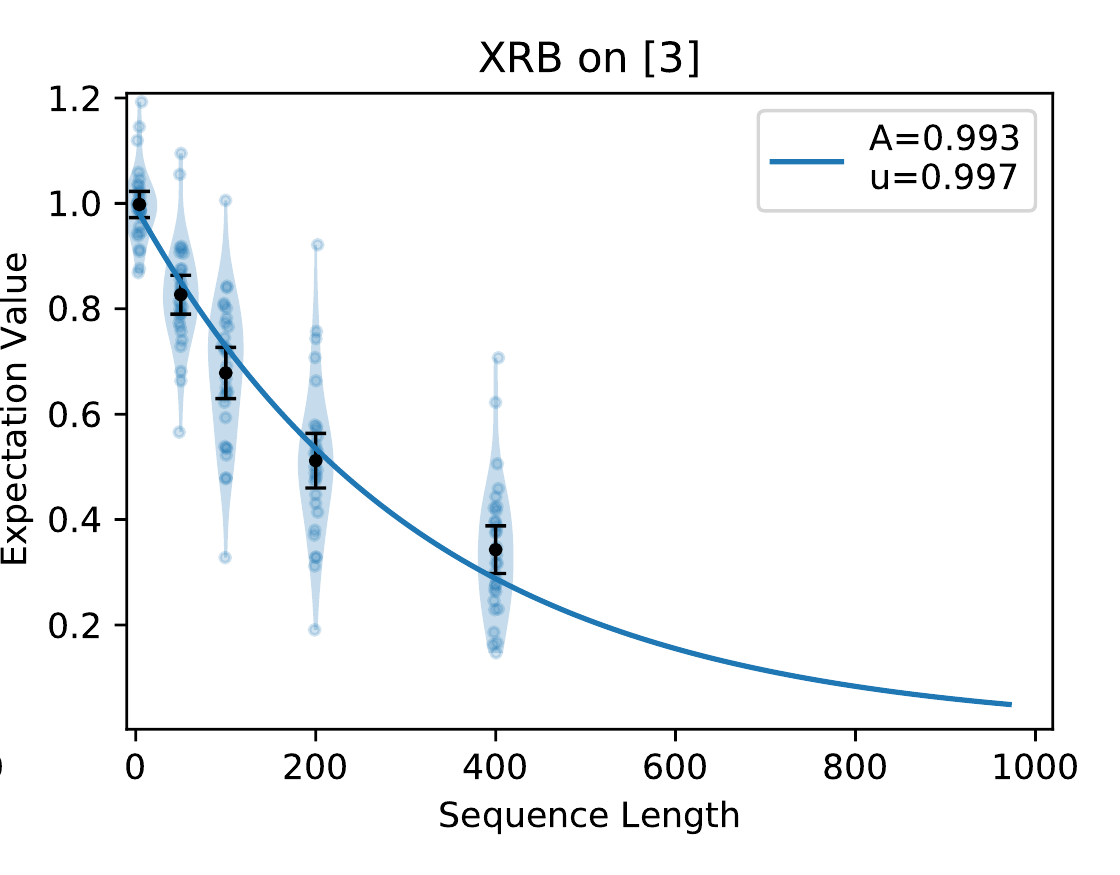}
\label{fig:1qxrb}}\\
\subfloat[Two-qubit Streamlined RB.]{\includegraphics[width=0.33\linewidth]{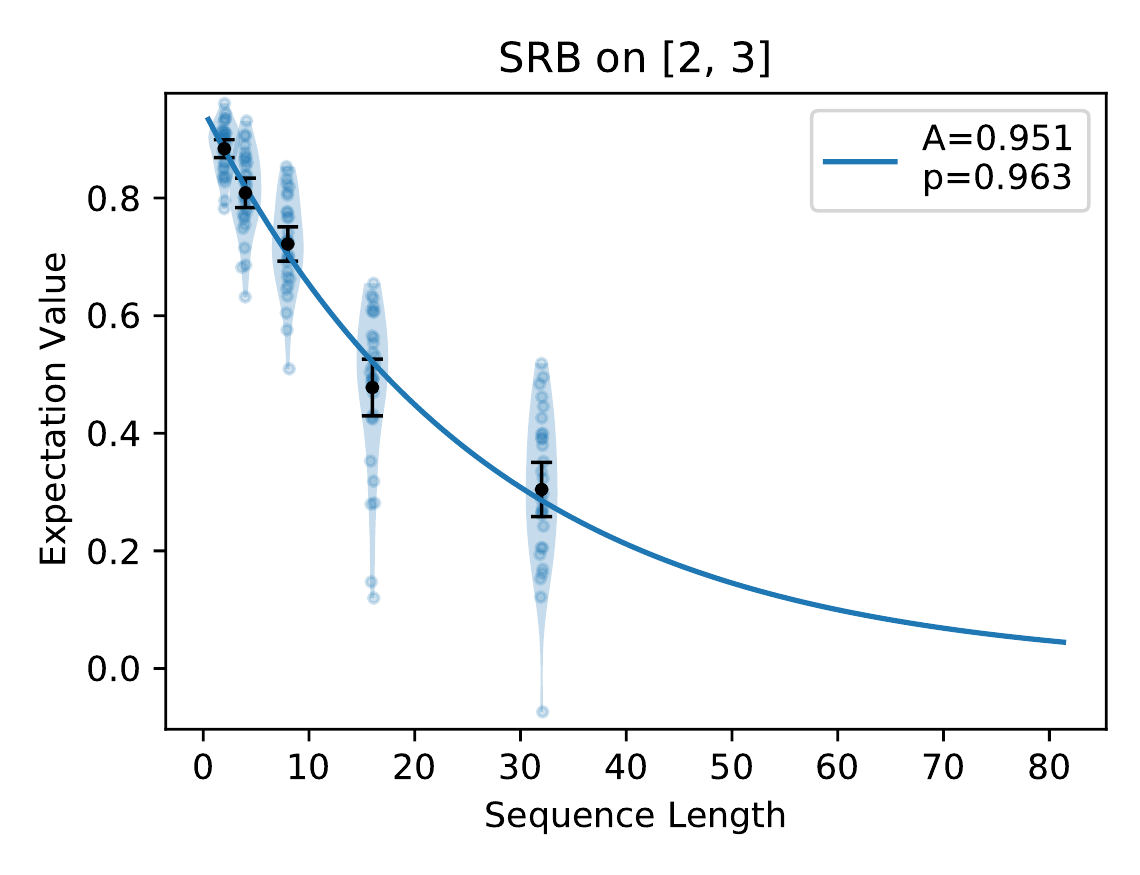}
\label{fig:2qsrb}}
\subfloat[Two-qubit Interleaved RB.]{\includegraphics[width=0.33\linewidth]{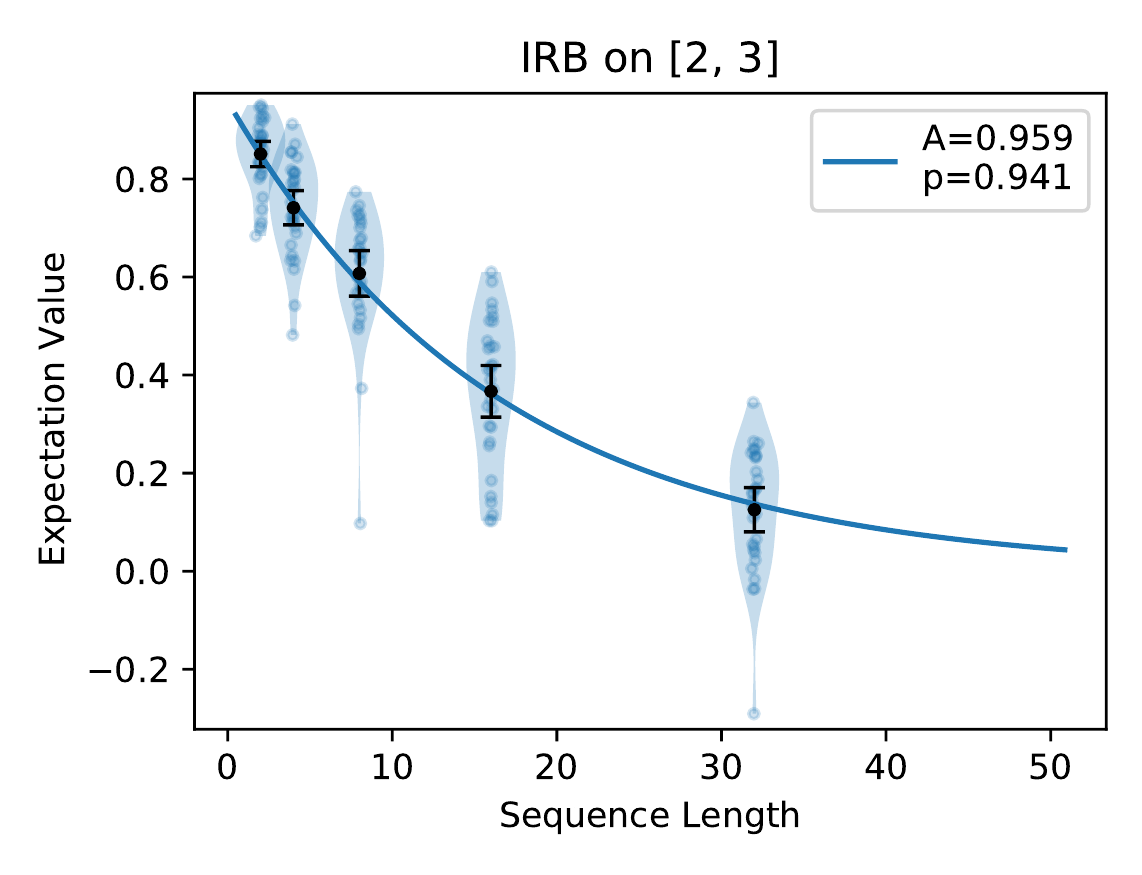}
\label{fig:2qirb}}
\subfloat[Two-qubit Extended RB.]{\includegraphics[width=0.33\linewidth]{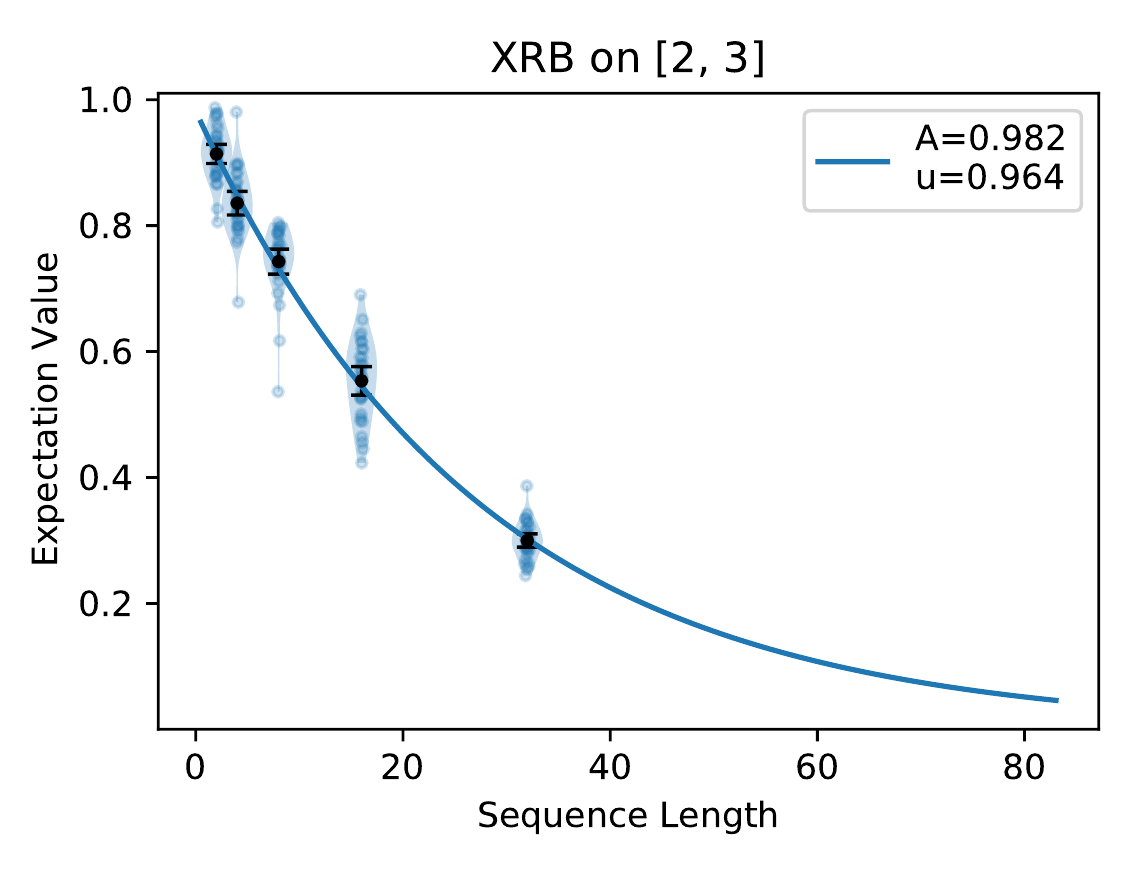}
\label{fig:2qxrb}}
\caption{RB results. We perform three kinds of RB protocols (SRB, IRB, XRB) on single qubit (Q3) and two qubits (Q2 and Q3) respectively. In each plot, each point corresponds to a measurement result of a random quantum circuit. The SRB and IRB curves are fitted with an exponential decay function $\mathrm{A}\mathrm{p}^\mathrm{m}$, while XRB curve is fitted with the function of $\mathrm{A}\mathrm{u}^\mathrm{m}$, where $\mathrm{p}$ or $\mathrm{u}$ is the decay parameter. $\mathrm{A}$ is a constant which encapsulates the common state preparation and readout errors. The sequence length $\mathrm{m}$ is expressed in terms of Clifford gates. The RB data collected with {\it QubiC} are passed to the True-Q software \cite{beale_stefanie_j_2020_3945250} to fit the data and extract the parameters.}
\label{fig:rb}
\end{figure}

In order to validate the performance of the automatic qubit characterization and gate optimization protocol, we implement randomized benchmarking (RB) \cite{knill2008randomized} sequences on an 8-qubit quantum information processor \cite{kreikebaum2020improving} at the Advanced Quantum Testbed (AQT) in Lawrence Berkeley National Laboratory (LBNL), and evaluated the single-qubit and two-qubit gate fidelities.
The streamlined randomized benchmarking (SRB) \cite{magesan2011scalable} is the standard protocol to characterize the probability of an error occurring during a gate, while the interleaved randomized benchmarking (IRB) \cite{magesan2012efficient} enhances SRB by estimating the average gate fidelity of specific gates.
The extended randomized benchmarking (XRB) \cite{wallman2015estimating} enables us to distinguish unitary errors from stochastic errors in combination with SRB. 
As an illustration, the single-qubit X(90) and two-qubit CNOT gates are optimized using {\it QubiC} automatic calibration protocol, and evaluated by SRB, IRB, and XRB, as shown in Fig.~\ref{fig:rb}.
Table~\ref{tab:infidelity} illustrates the single-qubit and two-qubit Clifford gate infidelities and errors, which demonstrate that the automatic calibration protocol has the capability to deliver high-fidelity gates on state-of-the-art processors.



\begin{table}
\begin{threeparttable}
  \caption{Gate infidelities and errors.}
  \label{tab:infidelity}
  \begin{tabular}{ccccc}
    \toprule
    Gate & Gate Infidelity \tnote{$\dag$} & Process Infidelity & Unitary Error & Stochastic Error \\
    \midrule
    Single-qubit Clifford & $4.9(1.1) \times 10^{-4}$ & $1.3(1) \times 10^{-3}$ & $1.9(9) \times 10^{-4}$ & $1.2(1) \times 10^{-3}$ \\
    Two-qubit Clifford & $1.4(3) \times 10^{-2}$ & $3.5(2) \times 10^{-2}$ & $1.7(2) \times 10^{-2}$ & $1.7(1) \times 10^{-2}$ \\
  \bottomrule
\end{tabular}
\begin{tablenotes}\footnotesize
\item[$\dag$] The gate infidelity was calculated as $\frac{3}{4}(1-\frac{p_{\mathrm{IRB}}}{p_{\mathrm{SRB}}})$, where $p_{\mathrm{IRB}}$ and $p_{\mathrm{SRB}}$ are IRB and SRB decay parameters of $\mathrm{A}\mathrm{p}^\mathrm{m}$ curves \cite{magesan2012efficient}.
\end{tablenotes}
\end{threeparttable}
\end{table}

\section{Conclusion}
We develop an efficient and systematic method to automatically characterize qubits and optimize gates with {\it QubiC}, which is a customized FPGA-based Qubit Control system at LBNL. 
For the single-qubit calibration, the mutli-dimensional loss-based optimization is proposed to characterize each qubit on the quantum processor.
With the stacking of consecutive identical gates, the drive amplitude is optimized to finely tune the single-qubit gate. 
For the two-qubit CNOT gate calibration, we stack multiple CR pulses and sweep CR pulse amplitude to achieve the full entanglement of the two qubits. 
We propose a full XY-plane measurement to project the qubit state to any angle on the XY-plane, and extract the CNOT parameters from curve fitting by only implementing one set of measurement.
The automatic qubit characterization and gate optimization protocols are validated by performing RB sequences on a superconducting quantum information processor.   
The measurement results demonstrate our method is capable of delivering high-fidelity gates on state-of-the-art processors.

\section*{Acknowledgement}
This work was supported by the Advanced Scientific Computing Research Testbeds for Science program, by the High Energy Physics QUANTISED program, and by the Quantum Systems Accelerator under the Office of Science of the U.S. Department of Energy under Contract No. DE-AC02-05CH11231.

\bibliographystyle{unsrt} 
\bibliography{references}

\end{document}